\documentclass[preprintnumbers,amsmath,amssymbm,preprint]{revtex4}
\usepackage{epsfig}
\usepackage{graphicx}

\begin{document}
\title{Rotating black holes can have 
short bristles}
\author{Shahar Hod}
\affiliation{The Ruppin Academic Center, Emeq Hefer 40250, Israel}
\affiliation{ } \affiliation{The Hadassah Institute, Jerusalem
91010, Israel}
\date{\today}

\begin{abstract}
The elegant `no short hair' theorem states that, if a
spherically-symmetric static black hole has hair, then this hair
must extend beyond 3/2 the horizon radius. In the present paper we
provide evidence for the failure of this theorem beyond the regime
of spherically-symmetric static black holes. In particular, we show
that rotating black holes can support extremely short-range
stationary scalar configurations (linearized scalar `clouds') in
their exterior regions. To that end, we solve analytically the
Klein-Gordon-Kerr-Newman wave equation for a {\it linearized}
massive scalar field in the regime of large scalar masses.
\end{abstract}
\bigskip
\maketitle

\section {Introduction.}

Within the framework of classical general relativity, the black-hole
horizon acts as a one-way membrane which irreversibly absorbs matter
fields and radiation. This remarkable property of the black-hole
horizon suggests, in particular, that static matter configurations
can not be supported in the spacetime region outside the black-hole
horizon. This expectation is nicely summarized in Wheeler's famous
dictum ``a black hole has no hair" \cite{Whee,Car}, which suggests
that the spacetime geometries of all asymptotically flat stationary
black holes are uniquely described by the three-parameter family
\cite{Notethr} of the Kerr-Newman electrovacuum solution
\cite{Chan,Kerr,Newman}.

The `no-hair' conjecture \cite{Whee,Car} has attracted much
attention over the years from both physicists and mathematicians.
Early investigations of the conjecture have ruled out the existence
of static hairy black-hole configurations made of scalar fields
\cite{Chas}, spinor fields \cite{Hart}, and massive vector fields
\cite{BekVec}. However, the early 90's have witnessed the discovery
of a variety of regular \cite{Noteregular} hairy black-hole
configurations, the first of which were the `colored' black holes
which are solutions of the coupled Einstein-Yang-Mills equations
\cite{BizCol}. It has soon been realized that many non-linear matter
fields \cite{Notenonl}, when coupled to the Einstein field
equations, can lead to the formation of hairy black-hole
configurations
\cite{Lavr,BizCham,Green,Stra,BiWa,EYMH,Volkov,BiCh,Lav1,Lav2,Bizw}.

The validity of the original no-hair conjecture \cite{Whee,Car} has
become highly doubtful since the discovery of these non-linear
\cite{BizCol,Lavr,BizCham,Green,Stra,BiWa,EYMH,Volkov,BiCh,Lav1,Lav2,Bizw}
hairy black-hole configurations \cite{Notevio}. The current
situation naturally gives rise to the following question: Is it
possible to formulate a more modest (and robust) alternative to the
original no hair conjecture?

A very intriguing attempt to reveal the generic characteristics of
hairy black-hole configurations was made in \cite{Nun}: A `{\it no
short hair}' theorem was proved, according to which static
spherically-symmetric black holes cannot support short hair. In
particular, it was shown in \cite{Nun} that, in all Einstein-matter
theories in which static hairy black-hole configurations have been
discovered, the effective length of the outside hair is bounded from
below by \cite{Notetheo}
\begin{equation}\label{Eq1}
r_{\text{hair}}>{3\over 2}r_{\text{H}}\  ,
\end{equation}
where $r_{\text{H}}$ is the horizon-radius of the black hole. This
`no short hair' theorem was suggested \cite{Nun} as an alternative
to the original \cite{Whee,Car} `no hair' conjecture.

It is worth emphasizing that the formal proof of the lower bound
(\ref{Eq1}) provided in \cite{Nun} is restricted to the static
sector of spherically-symmetric black holes. Nevertheless, it was
conjectured \cite{Nun} that the `no short hair' bound (\ref{Eq1})
can be generalized in the form
\begin{equation}\label{Eq2}
r_{\text{hair}}>{3\over 2}\sqrt{{A_{\text{H}}}\over{4\pi}}\ \ \ \ ;
\ \ \ \ A_{\text{H}}\equiv {\text{horizon area}}\
\end{equation}
to include the cases of non spherically symmetric stationary hairy
black-hole configurations.

The main goal of the present paper is to test the validity of the
`no short hair' conjecture beyond the regime of spherically
symmetric static black holes. In particular, we shall explore here
the physical properties of non spherically symmetric rotating black
holes coupled to linearized stationary (rather than static) scalar
matter configurations. (It should be emphasized that the scalar
fields we consider have a time dependence of the form $e^{-i\omega
t}$ [see Eq. (\ref{Eq10}) below]. However, physical quantities, like
the energy-momentum tensor itself, are time-{\it independent}).

\section{Composed black-hole-scalar-field configurations.}

While early no hair theorems have shown that asymptotically flat
black holes cannot support regular static scalar configurations in
their exterior regions \cite{Chas}, they have not ruled out the
existence of non-static composed black-hole-scalar-field
configurations. In fact, it has recently \cite{Hodstat} been
demonstrated that rotating black holes can support linearized
stationary scalar configurations (scalar `clouds'
\cite{Notecloud,Barr}) in their exterior regions. Since non-linear
(self-interaction) effects tend to stabilize the outside hair
\cite{Nun,Hod11}, we conjectured in \cite{Hodstat} the existence of
rotating black hole solutions endowed with genuine non-static scalar
hair. These non-static hairy black-hole-scalar-field configurations
are the non-linear counterparts of the linear scalar clouds studied
analytically in \cite{Hodstat}. In a very interesting Letter,
Herdeiro and Radu \cite{HerRa} have recently solved numerically the
non-linear coupled Einstein-scalar equations, and confirmed the
existence of these non-static hairy black-hole configurations.

The composed black-hole-scalar-field configurations \cite{Notebos}
explored in \cite{Hodstat,HerRa} are intimately related to the
intriguing phenomenon of superradiant scattering of bosonic fields
in rotating black-hole spacetimes \cite{Zel,PressTeu1,Ins1,Ins2}. In
particular, the linearized stationary scalar configurations studied
in \cite{Hodstat,HerRa} are characterized by orbital frequencies
which are integer multiples of the central black-hole angular
frequency \cite{Noteunits}:
\begin{equation}\label{Eq3}
\omega_{\text{field}}=m\Omega_{\text{H}}\ \ \ \ \text{with}\ \ \ \
m=1,2,3,...\  .
\end{equation}

It is well-established \cite{Zel,PressTeu1,Ins1,Ins2} that the
energy flux of the field into the central spinning black hole
vanishes for bosonic modes which satisfy the relation (\ref{Eq3}).
In this case, the bosonic field is not swallowed by the central
black hole. This suggests that stationary bosonic configurations
which are in resonance with the central spinning black hole (that
is, bosonic fields with orbital frequencies
$\omega_{\text{field}}=m\Omega_{\text{H}}$) may survive in the
spacetime region exterior to the black-hole horizon.

In order to have genuine stationary (non-decaying) field
configurations around the central black hole, one should also
prevent the field from escaping to infinity. A natural confinement
mechanism is provided by the gravitational attraction between the
massive field and the central black hole. In particular, for a
scalar field of mass $\mu$, low frequency field modes in the regime
\cite{Notedim}
\begin{equation}\label{Eq4}
\omega^2<\mu^2
\end{equation}
are confined to the vicinity of the central black hole.

As discussed above, the main goal of the present paper is to test
the validity of the `no short hair' conjecture (\ref{Eq1})
\cite{Nun} beyond the regime of spherically-symmetric static black
holes. To that end, we shall analyze the physical properties of the
non-static (rotating) black-hole-scalar-field configurations
\cite{Hodstat,HerRa} in the eikonal regime
\begin{equation}\label{Eq5}
M\mu\gg1\  ,
\end{equation}
where $M$ is the mass of the central spinning black hole.

\section{Description of the system.}

The physical system we consider consists of a massive scalar field
$\Psi$ linearly coupled \cite{Notelin} to an extremal Kerr-Newman
black hole of mass $M$, angular-momentum per unit mass $a$, and
electric charge $Q$. In Boyer-Lindquist coordinates
$(t,r,\theta,\phi)$ the spacetime metric is given by
\cite{Chan,Kerr,Newman}
\begin{eqnarray}\label{Eq6}
ds^2=-{{\Delta}\over{\rho^2}}(dt-a\sin^2\theta
d\phi)^2+{{\rho^2}\over{\Delta}}dr^2+\rho^2
d\theta^2+{{\sin^2\theta}\over{\rho^2}}\big[a
dt-(r^2+a^2)d\phi\big]^2
\end{eqnarray}
where $\Delta\equiv r^2-2Mr+a^2+Q^2$ and $\rho\equiv
r^2+a^2\cos^2\theta$. The extremality condition implies that the
degenerate horizon of the black hole is located at
\begin{equation}\label{Eq7}
r_{\text{H}}=M=\sqrt{a^2+Q^2}\  .
\end{equation}
The angular velocity of the black hole is given by
\cite{Chan,Kerr,Newman}
\begin{equation}\label{Eq8}
\Omega_{\text{H}}={{a}\over{M^2+a^2}}\  .
\end{equation}

The dynamics of the linearized massive scalar field $\Psi$ in the
Kerr-Newman black-hole spacetime is governed by the Klein-Gordon
(Teukolsky) wave equation
\begin{equation}\label{Eq9}
(\nabla^\nu\nabla_{\nu}-\mu^2)\Psi=0\  .
\end{equation}
It proves useful to use the ansatz \cite{Noteanz}
\begin{equation}\label{Eq10}
\Psi(t,r,\theta,\phi)=\int\sum_{l,m}e^{im\phi}{S_{lm}}(\theta;{s}\epsilon){R_{lm}}(r;s,\mu,\omega)e^{-i\omega
t}d\omega\
\end{equation}
for the scalar wave field in (\ref{Eq9}), where
\begin{equation}\label{Eq11}
{s}\equiv {{a}\over{M}}
\end{equation}
is the dimensionless angular-momentum (spin) of the black hole, and
\begin{equation}\label{Eq12}
\epsilon\equiv M\sqrt{\mu^2-\omega^2}\ .
\end{equation}

The angular equation for ${S_{lm}}(\theta;{s}\epsilon)$, which is
obtained from the substitution of (\ref{Eq10}) into (\ref{Eq9}), is
given by \cite{Stro,Heun,Fiz1,Teuk,Abram,Hodasy}
\begin{eqnarray}\label{Eq13}
{1\over {\sin\theta}}{{d}\over{\theta}}\Big(\sin\theta {{d
S_{lm}}\over{d\theta}}\Big)
+\Big[K_{lm}+({s}\epsilon)^2\sin^2\theta-{{m^2}\over{\sin^2\theta}}\Big]S_{lm}=0\
.
\end{eqnarray}
This angular equation is supplemented by the requirement that the
angular functions $S_{lm}(\theta;{s}\epsilon)$ \cite{Noteangu} be
regular at the poles $\theta=0$ and $\theta=\pi$. These boundary
conditions single out the discrete set of angular eigenvalues
$\{K_{lm}({s}\epsilon)\}$ with $l\geq |m|$ \cite{Abram}. We shall
henceforth consider equatorial scalar modes in the eikonal regime
\begin{equation}\label{Eq14}
l=m\gg1\ \ \ \ \text{and}\ \ \ \ {s}\epsilon\gg 1\  ,
\end{equation}
in which case the angular eigenvalues are given by
\cite{Yang,Notetbp}
\begin{equation}\label{Eq15}
K_{mm}({s}\epsilon)=m^2-({s}\epsilon)^2+O(m)\  .
\end{equation}

The radial equation for ${R_{lm}}$, which is obtained from the
substitution of (\ref{Eq10}) into (\ref{Eq9}), is given by
\cite{Teuk,Stro}
\begin{equation}\label{Eq16}
\Delta{{d}
\over{dr}}\Big(\Delta{{dR_{lm}}\over{dr}}\Big)+\Big[[(r^2+a^2)\omega-ma]^2
+\Delta[2ma\omega-\mu^2(r^2+a^2)-K_{lm}]\Big]R_{lm}=0\ .
\end{equation}
Note that the radial equation (\ref{Eq16}) for ${R_{lm}}$ is coupled
to the angular equation (\ref{Eq13}) for ${S_{lm}}$ through the
angular eigenvalues $\{{K_{lm}}({s}\epsilon)\}$ \cite{Notebr}.

\section{Stationary bound-state resonances of the composed
black-hole-scalar-field system.}

In the present paper we shall explore the physical properties of the
linearized {\it stationary} scalar configurations which characterize
the composed Kerr-Newman-scalar-field system. These stationary
bound-state resonances of the bosonic field are characterized by the
critical frequency
\begin{equation}\label{Eq17}
\omega_{\text{field}}=\omega_{\text{c}}\equiv m\Omega_{\text{H}}\
\end{equation}
for superradiant scattering in the black-hole spacetime [see Eq.
(\ref{Eq3})]

The bound-state solutions of the radial equation (\ref{Eq16}) are
characterized by a decaying field at spatial infinity \cite{Ins2}:
\begin{equation}\label{Eq18}
R(r\to\infty)\sim {{1}\over{r}}e^{-\epsilon r/r_{\text{H}}}\
\end{equation}
with $\epsilon^2>0$ \cite{Noteepp}. Regular (finite energy) field
configurations are also bounded at the black-hole horizon:
\begin{equation}\label{Eq19}
R(r=r_{\text{H}})<\infty\  .
\end{equation}
The boundary conditions (\ref{Eq18}) and (\ref{Eq19}) single out the
discrete family of radial eigenfunctions [along with the associated
eigen field-masses, see Eq. (\ref{Eq31}) below] which characterize
the bound-state stationary scalar configurations.

We shall first obtain a simple analytic formula for the discrete
spectrum of field masses, $\{\mu(m,{s};n)\}$ \cite{Notenr}, which
characterize the stationary bound-state resonances of the massive
scalar fields in the extremal Kerr-Newman black-hole spacetime. To
that end, it proves useful to define a new dimensionless radial
coordinate \cite{Teuk,Stro}
\begin{equation}\label{Eq20}
x\equiv {{r-M}\over {M}}\  ,
\end{equation}
in terms of which the radial equation (\ref{Eq16}) becomes
\begin{equation}\label{Eq21}
x^2{{d^2R}\over{dx^2}}+2x{{dR}\over{dx}}+VR=0\  ,
\end{equation}
where $V\equiv
[M\omega_{\text{c}}(x+2)]^2-K+2Mm{s}\omega_{\text{c}}-(M\mu)^2[(x+1)^2+{s}^2]$.
Remarkably, this radial equation for $R(x)$ can be solved {\it
analytically} \cite{Abram,Hodstat}:
\begin{equation}\label{Eq22}
R(x)=C_1\times x^{-{1\over 2}+\beta}e^{-\epsilon x}M({1\over
2}+\beta-\kappa,1+2\beta,2\epsilon x)+C_2\times(\beta\to -\beta)\  ,
\end{equation}
where $M(a,b,z)$ is the confluent hypergeometric function
\cite{Abram} and $\{C_1,C_2\}$ are normalization constants. Here
\begin{equation}\label{Eq23}
\kappa\equiv{{\alpha}\over{\epsilon}}-\epsilon\ \ \ \ \text{with}\ \
\ \ \alpha\equiv
(M\omega_{\text{c}})^2={{(m{s})^2}\over{(1+{s}^2)^2}}\ ,
\end{equation}
and \cite{Notedel}
\begin{equation}\label{Eq24}
\beta^2\equiv{K+{1\over
4}-2Mm{s}\omega_{\text{c}}-(2M\omega_{\text{c}})^2+(M\mu)^2(1+{s}^2)}\
.
\end{equation}
The notation $(\beta\to -\beta)$ in (\ref{Eq22}) means ``replace
$\beta$ by $-\beta$ in the preceding term.". Taking cognizance of
Eqs. (\ref{Eq8}), (\ref{Eq15}) and (\ref{Eq17}), one can express
$\beta$ in the eikonal regime (\ref{Eq14}) in the form
\begin{equation}\label{Eq25}
\beta=\sqrt{\beta^2_0+\epsilon^2}\ \ \ \ \ \text{with}\ \ \ \ \
\beta^2_0\equiv m^2{{1-3{s}^2}\over{(1+{s}^2)^2}}[1+O(m^{-1})]\ .
\end{equation}

We shall now analyze the spatial behavior of the radial wave
function (\ref{Eq22}) in the asymptotic regimes $x\to 0$ and
$x\to\infty$:
\newline
(1) The behavior of the radial function (\ref{Eq22}) in the
near-horizon $x\ll 1$ region is given by \cite{Abram}
\begin{equation}\label{Eq26}
R(x\to 0)\to C_1\times x^{-{1\over 2}+\beta}+C_2\times x^{-{1\over
2}-\beta}\  .
\end{equation}
From Eq. (\ref{Eq26}) one learns that a well-behaved [see Eq.
(\ref{Eq19})] stationary field configuration is characterized by
\cite{Noteregu,Dolh}
\begin{equation}\label{Eq27}
C_2=0 \ \ \ \ \text{and}\ \ \ \ \Re\beta\geq {1\over 2}\  .
\end{equation}
\newline
(2) The behavior of the radial function (\ref{Eq22}) in the
asymptotic $x\to\infty$ region is given by \cite{Abram}
\begin{eqnarray}\label{Eq28}
R(x\to\infty)&\to& C_1\times(2\epsilon)^{\kappa-{1\over
2}-\beta}{{\Gamma(1+2\beta)}\over{\Gamma({1\over
2}+\beta+\kappa)}}x^{-1+\kappa}(-1)^{-{1\over
2}-\beta+\kappa}e^{-\epsilon x} \nonumber \\&& +
C_1\times(2\epsilon)^{-\kappa-{1\over
2}-\beta}{{\Gamma(1+2\beta)}\over{\Gamma({1\over
2}+\beta-\kappa)}}x^{-1-\kappa}e^{\epsilon x}\ .
\end{eqnarray}
The bound-state (finite-energy) scalar configurations are
characterized by asymptotically decaying eigenfunctions at large
distances from the central black hole [see Eq. (\ref{Eq18})]. Thus,
the coefficient of the growing exponent $e^{\epsilon x}$ in
(\ref{Eq28}) must be identically zero. This boundary condition
yields the resonance condition \cite{Noteabr}
\begin{equation}\label{Eq29}
{1\over 2}+\beta-\kappa=-n\ \ \ \ \text{with}\ \ \ n=0,1,2,...\  .
\end{equation}
for the linearized stationary bound-state resonances of the massive
scalar fields in the rotating Kerr-Newman black-hole spacetime.

Taking cognizance of Eqs. (\ref{Eq23}) and (\ref{Eq25}), one can
express the resonance condition (\ref{Eq29}) in the form
$\sqrt{\beta^2_0+\epsilon^2}=\alpha\epsilon^{-1}-\epsilon-(n+1/2)$,
which in the eikonal regime ($m\gg n+1/2$) yields the simple
relation [see Eqs. (\ref{Eq23}) and (\ref{Eq25})]
\begin{equation}\label{Eq30}
\epsilon=m{{{s}^2}\over{(1+{s}^2)\sqrt{1-{s}^2}}}[1+O(m^{-1})]\
\end{equation}
for the bound-state resonances in the regime
$0<{s}<{{1}\over{\sqrt{2}}}$. Finally, taking cognizance of the
relation (\ref{Eq12}), one finds
\begin{equation}\label{Eq31}
M\mu(m,{s})=m{{{s}}\over{(1+{s}^2)\sqrt{1-{s}^2}}}[1+O(m^{-1})]
\end{equation}
for the scalar field-masses which characterize the stationary
bound-state resonances of the composed Kerr-Newman-scalar-field
system.

\section{Effective lengths of the stationary bound-state scalar
configurations.}

Motivated by the intriguing `no short hair' theorem (\ref{Eq1})
\cite{Nun}, we shall now analyze the effective lengths of the
linearized stationary bound-state scalar configurations. Taking
cognizance of Eqs. (\ref{Eq27}) and (\ref{Eq29}), one can write the
radial function (\ref{Eq22}) for the stationary bound-state
configurations in the compact form
$R(x)=Ax^{-{1\over 2}+\beta}e^{-\epsilon x}L^{(2\beta)}_n(2\epsilon
x)$,
where $A$ is a normalization constant and $L^{(2\beta)}_n(x)$ are
the generalized Laguerre Polynomials \cite{Notesea}. In particular,
the fundamental ($n=0$) bound-state resonance
is characterized by the remarkably simple radial eigenfunction
\cite{Noten0}
\begin{equation}\label{Eq32}
R^{(0)}(x)=Ax^{-{1\over 2}+\beta}e^{-\epsilon x}\  .
\end{equation}
The radial distribution (\ref{Eq32}) peaks at
$x_{\text{peak}}=(\beta-1/2)/\epsilon$, which implies [see Eqs.
(\ref{Eq25}) and (\ref{Eq30})]
\begin{equation}\label{Eq33}
x_{\text{peak}}={{1-2{s}^2}\over{{s}^2}}[1+O(m^{-1})]\  .
\end{equation}

Equation (\ref{Eq33}) reveals the remarkable fact that, the
bound-state stationary scalar configurations can be made arbitrarily
compact. In particular, one finds
\begin{equation}\label{Eq34}
x_{\text{peak}}\to 0\ \ \ \text{for}\ \ \ {s}\to
{{1}\over{\sqrt{2}}}\  .
\end{equation}
One therefore concludes that rotating black holes can support
extremely short-range stationary scalar configurations (linearized
scalar `clouds') in their exterior regions. Our analysis thus
provides evidence for the failure of the `no short hair' theorem
\cite{Nun} beyond the regime of spherically-symmetric static black
holes.

\section{Summary.}

In a very intriguing Letter \cite{Nun} a remarkable observation was
made according to which static spherically-symmetric black holes
cannot have short hair. In particular, it was proved \cite{Nun} that
if a spherically-symmetric static black hole has hair, then this
hair must extend beyond $3/2$ the horizon radius [see Eq.
(\ref{Eq1})]. The main goal of the present paper was to test the
general validity of this `no short hair' conjecture.

To that end, we have analyzed the physical properties of non
spherically symmetric rotating black holes coupled to stationary
(rather than static) linear matter configurations. In particular, we
have shown that rotating Kerr-Newman black holes can support
extremely {\it short}-range stationary scalar configurations
(linearized scalar `bristles') in their exterior regions. Our
analysis thus provides compelling evidence for the failure of the
`no short hair' conjecture (\ref{Eq1}) \cite{Nun} beyond the regime
of spherically-symmetric static black holes.

\bigskip
\noindent
{\bf ACKNOWLEDGMENTS}
\bigskip

This research is supported by the Carmel Science Foundation. I thank
C. A. R. Herdeiro and E. Radu for helpful correspondence. I would
also like to thank Yael Oren, Arbel M. Ongo and Ayelet B. Lata for
stimulating discussions.



\begin{thebibliography}{99}
\bibitem{Notethr} The three physical parameters characterizing
the black-hole spacetime are the mass, charge, and angular momentum
of the black hole, see Eq. (\ref{Eq6}) below.
\bibitem{Noteregular} That is, the matter fields are well behaved on and
outside the black-hole horizon.
\bibitem{Notenonl} That is, matter fields with (non-linear) self-interaction terms.
\bibitem{Notevio} It is worth noting, however, that most of these
hairy black-hole configurations are dynamically unstable.
\bibitem{Notetheo} It is worth emphasizing that, within the sector of spherically symmetric
black holes, this lower bound is universal in the sense that it is
parameter and theory independent \cite{Nun}.
\bibitem{Notecloud} We use here the terminology of \cite{Barr} to reflect
the fact that these stationary bound-state configurations are made
of linearized scalar fields.
\bibitem{Notebos} It is expected that non-static hairy black-hole
configurations exist for other bosonic fields as well, see also
\cite{HerRa}.
\bibitem{Noteunits} We shall use natural units in which
$G=c=\hbar=1$.
\bibitem{Notedim} Note that $\mu$ stands for $\mu/\hbar$.
Thus, the field parameter $\mu$ has the dimensions of
(length)$^{-1}$.
\bibitem{Noteanz} Here $\omega,\ l,$ and $m$ are respectively the conserved
frequency, the spheroidal harmonic index, and the azimuthal harmonic
index of the wave mode [see Eq. (\ref{Eq13}) below].
\bibitem{Noteangu} The angular functions $S_{lm}(\theta;{s}\epsilon)$ are known as
the spheroidal harmonic functions. In the ${s}\epsilon\ll 1$ limit
these angular functions become the familiar spherical harmonics.
\bibitem{Notebr} We shall henceforth omit the harmonic indexes $l$ and $m$ for brevity.
\bibitem{Noteepp} One may choose $\epsilon>0$ without loss of generality.
\bibitem{Notedel} One may choose $\Re\beta>0$ without loss of generality.
\bibitem{Noteabr} Here we have used equation 6.1.7 of \cite{Abram}.
\bibitem{Noten0} Here we have used the identity $L^{(2\beta)}_0(x)\equiv
1$.
\end{thebibliography}
\end{document}